\documentclass[iop]{emulateapj}

\usepackage{natbib,graphicx}
\begin{document}

\title{The 2011 Outburst of Recurrent Nova T~Pyx:\\
 X-ray Observations Expose the White Dwarf Mass and Ejection Dynamics}

\author{Laura~Chomiuk\altaffilmark{1,2}, Thomas~Nelson\altaffilmark{3}, Koji~Mukai\altaffilmark{4,5}, J.~L.~Sokoloski\altaffilmark{6}, Michael~P.~Rupen\altaffilmark{2}, Kim~L.~Page\altaffilmark{7}, Julian~P.~Osborne\altaffilmark{7}, Erik Kuulkers\altaffilmark{8}, Amy~J.~Mioduszewski\altaffilmark{2}, Nirupam~Roy\altaffilmark{9}, Jennifer~Weston\altaffilmark{6}, \& Miriam~I.~Krauss\altaffilmark{2}}

\altaffiltext{1}{Department of Physics and Astronomy, Michigan State University, East Lansing, MI 48824, USA}
\altaffiltext{2}{National Radio Astronomy Observatory, P.O. Box O, Socorro, NM 87801, USA}
\altaffiltext{3}{School of Physics and Astronomy, University of Minnesota, 115 Church St SE, Minneapolis, MN 55455, USA}
\altaffiltext{4}{CRESST and X-ray Astrophysics Laboratory, NASA/GSFC, Greenbelt, MD 20771, USA}
\altaffiltext{5}{Center for Space Science and Technology, University of Maryland Baltimore County, 1000 Hilltop Circle, Baltimore MD 21250, USA}
\altaffiltext{6}{Columbia Astrophysics Laboratory, Columbia University, New York, NY, USA}
\altaffiltext{7}{Department of Physics and Astronomy, University of Leicester, Leicester, LE1 7RH, UK}
\altaffiltext{8}{European Space Astronomy Centre (ESA/ESAC), Science Operations Department, 28691 Villanueva de la Ca–ada, Madrid, Spain}
\altaffiltext{9}{Max Planck Institut f\"{u}r Radioastronomie, Auf dem H\"{u}gel 69 53121 Bonn, Germany}
\email{chomiuk@pa.msu.edu}

\begin{abstract}
The recurrent nova T~Pyx underwent its sixth historical outburst in 2011, and became the
subject of an intensive multi-wavelength observational campaign. We analyze data from the
\emph{Swift} and \emph{Suzaku} satellites to produce a detailed X-ray light curve augmented by
epochs of spectral information. X-ray observations yield mostly non-detections in the first four
months of outburst, but both a super-soft and hard X-ray component rise rapidly after Day 115. The
super-soft X-ray component, attributable to the photosphere of the nuclear-burning white dwarf, is
relatively cool ($\sim$45 eV) and implies that the white dwarf in T~Pyx is significantly below the
Chandrasekhar mass ($\sim$1 M$_{\odot}$). The late turn-on time of the super-soft component yields
a large nova ejecta mass ($\gtrsim10^{-5}$ M$_{\odot}$), consistent with estimates at other wavelengths.  The hard X-ray component
is well fit by a $\sim$1 keV thermal plasma, and is attributed to shocks internal
to the 2011 nova ejecta.
 The presence of a strong oxygen line in this thermal plasma on Day 194 requires a significantly super-solar abundance of
oxygen and implies that the ejecta are polluted by white dwarf material.
The X-ray light curve can be explained by a dual-phase ejection,
with a significant delay between the first and second ejection phases, and the second ejection
finally released two months after outburst. A delayed ejection is consistent with optical and
radio observations of T~Pyx, but the physical mechanism producing such a delay remains a mystery.
\end{abstract}


\section{Introduction}\label{intro}

The five thermonuclear explosions of T~Pyxidis observed in 1890, 1902, 1920, 1944 and 1966 earned the system its place as the prototypical recurrent nova, but have also highlighted our poor understanding of many aspects of binary evolution and nova theory. The community has waited anxiously for the next outburst of T~Pyx in order to study this peculiar system with modern multi-wavelength capabilities, and T~Pyx finally obliged by entering its sixth recorded outburst in April 2011. High-quality panchromatic observations are now revealing a host of new surprises for this system.

A nova is a transient event marking a thermonuclear runaway on the surface of an accreting white dwarf. The white dwarf accretes hydrogen-rich material from a companion star, and this accreted material settles down into a thin degenerate layer on the surface of the white dwarf. The pressure and temperature in this layer increase until explosive nuclear burning begins, and the bulk of the accreted envelope is expelled from the white dwarf at hundreds to thousands of km s$^{-1}$.

Novae are expected to recur on an accreting white dwarf with a timescale primarily determined by the white dwarf mass and accretion rate \citep[e.g.,][]{Yaron_etal05, Wolf13}. Predicted recurrence timescales vary widely ($\sim$1--10$^{8}$ years; \citealt{Yaron_etal05}), and those novae repeating on historical timescales have been dubbed ``recurrent" novae. Theoretically, we expect recurrent novae to occur in binaries where massive white dwarfs accrete at high rates, because more massive white dwarfs have higher surface gravities, meaning that the critical conditions are reached for smaller accreted envelopes, and higher accretion rate systems accrue this trigger mass in less time. 

Even before 2011, T~Pyx flew in the face of our expectations for recurrent novae. The evolution of its optical light curve is slow, showing a several months-long plateau around maximum light and a relatively slow decline from this maximum \citep{Schaefer10b}. T~Pyx has a short orbital period (1.83 hours; \citealt{Uthas10}), solidly below the cataclysmic-variable (CV) period gap. According to the theory of CV evolution, such short-period systems should have, on average, very low accretion rates \citep[e.g.,][]{Knigge11a}, but observations in quiescence---and the short nova recurrence time---imply that T~Pyx has an accretion rate orders of magnitude higher than these expectations \citep{Gilmozzi_Selvelli07, Selvelli08}. In addition, measured binary parameters imply that the white dwarf in T~Pyx may be significantly less massive than the Chandrasekhar mass \citep{Uthas10}, in contrast with common assumptions for recurrent novae.

The high accretion rate in T~Pyx does not appear to be sustainable, as it is exceeds expectations by several orders of magnitude for mass transfer rates driven by gravitational radiation (the commonly-accepted mass transfer mechanism at such short orbital periods). \citet{Knigge00} and \citet{Schaefer10a} have hypothesized that mass transfer in T~Pyx is in a short-term elevated state, perhaps incited by a powerful nova outburst which occurred during the 1800s (before regular records were kept on T~Pyx). Before this postulated event, T~Pyx may have been a typical CV below the period gap, with a very low accretion rate and long intervals between novae. However, after the hypothesized explosion, the hot white dwarf irradiated the companion star and induced an unusually high mass-transfer rate. Perhaps this irradiation power is slowly dwindling and the accretion rate is gradually declining, explaining the increasing intervals between nova events observed for T~Pyx throughout the last century. Observational tests of this hypothesis have reached divergent conclusions as to whether there is evidence for a secular decline in T~Pyx's accretion rate \citep[e.g.,][]{Schaefer13, Godon14}.

\begin{figure*}[t]
\begin{center}
\includegraphics[height=6.8in]{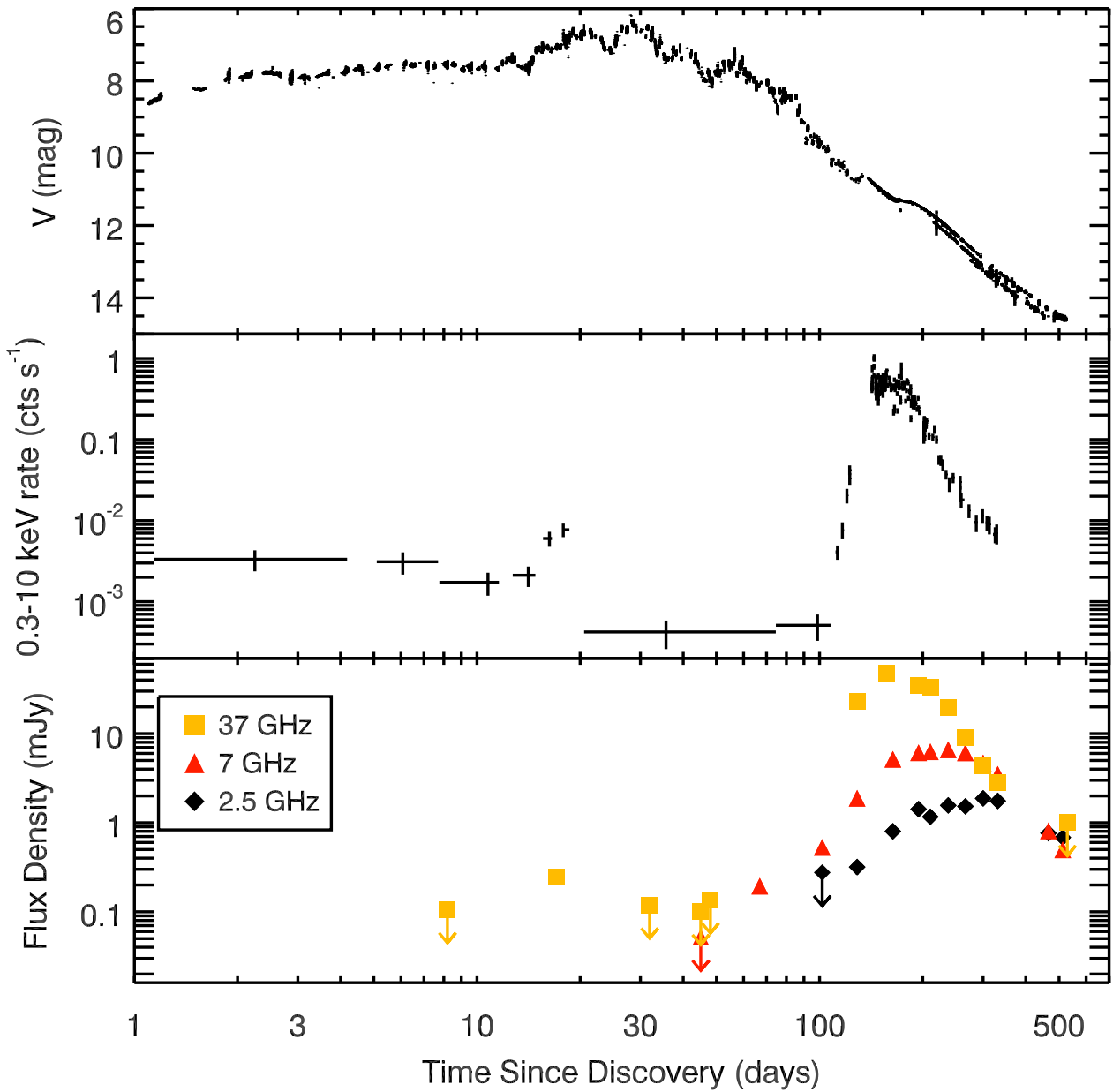}
\caption{Overview of the 2011 outburst of T~Pyx at optical, X-ray, and radio wavelengths.  {\it Top:} V-band optical data from the AAVSO. {\it Middle:} 0.3--10 keV X-ray light curve obtained with {\it Swift}/XRT.  {\it Bottom: } VLA radio light curve at 2.5, 7, and 37 GHz (Paper I). }
\label{rxo}
\end{center}
\end{figure*}

Regardless of T~Pyx's history, it is clear that T~Pyx provides an opportunity to test an unusual corner of nova parameter space. Compared with other novae, we have a thorough understanding of the binary system's parameters and the accretion rate \citep{Selvelli08, Uthas10}. With multi-wavelength data collected from the 2011 outburst, we can measure key properties of the nova event, like ejected mass, and compare them with predictions from nova models. With this goal in mind, campaigns have been carried out across the entire electromagnetic spectrum, providing an exquisitely detailed picture of the 2011 nova outburst of T~Pyx \citep{Chesneau11, Kuulkers11b, Kuulkers11a, Shore11, Shore13, Evans12, Imamura_Tanabe12, Nelson12t, Ederoclite13, Williams13, Schaefer13, Patterson13, Sokoloski13, Tofflemire13, DeGennaroAquino14, Godon14, Surina14}.

At every wavelength studied so far, the 2011 outburst of T~Pyx shows surprising features when compared to expectations for recurrent novae. As in previous outbursts, the optical light curve shows a sort of plateau for three months (Figure \ref{rxo}), implying that the optical photosphere is roughly constant in size for $\sim$90 days after thermonuclear runaway \citep{Shore13}. Optical spectroscopy shows that T~Pyx is an unusual ``hyper-hybrid" nova, switching from He/N class to \ion{Fe}{2} class around Day 10, and then back to He/N on Day 65 \citep{Williams12, Williams13, Ederoclite13, Surina14}. The radio light curve and optical measurements of the change in the binary period after the nova event imply a large ejected mass ($\sim$10$^{-4}-10^{-5}$ M$_{\odot}$; \citealt{Nelson12t} [henceforth Paper I], \citealt{Patterson13}), rather than the $\sim$10$^{-6}-10^{-7}$ M$_{\odot}$ expected for recurrent novae. Radio light curves also show a late and steep rise (Figure \ref{rxo}; Paper I), implying that either the ejecta in T~Pyx were very cold ($<$200 K) during the first $\sim$50 days of the outburst, or the bulk of the nova ejecta stalled at an $\sim$AU-scale radius until it was finally expelled $\sim$50 days after the thermonuclear runaway (such a delay might also explain the long plateau in the optical light curve; \citealt{Shore13}).

In this work, we focus on X-ray observations of the 2011 outburst of T~Pyx obtained with {\it Swift} and {\it Suzaku}, and compare our results with inferences from other wavelengths. X-ray emission from novae can be split into two broad classes, which may, but need not, exist contemporaneously \citep{Krautter08}. The first class is super-soft X-ray emission, characterized by effective temperatures between 10$^{5}$ and 10$^{6}$ K, and luminosities in the range 10$^{36}$--10$^{38}$ erg s$^{-1}$.  High resolution spectra obtained with the grating instruments onboard {\it Chandra} and {\it XMM-Newton} have confirmed that this emission originates near the white dwarf photosphere \citep{Nelson08, Rauch10, Ness11, Orio12}.  In some novae (e.g., V2491 Cyg, RS Oph) the soft X-ray flux is continuum emission that most likely originates at the white dwarf photosphere.  In other cases, the soft X-rays are associated with strong lines of H and He-like carbon, nitrogen and oxygen that likely indicate scattered photospheric emission \citep[see ][and references therein]{Ness13}.  Super-soft X-ray emission only becomes visible at later stages of the nova outburst, once the ejecta have become optically thin to the radiation from the hot, still burning white-dwarf surface layers; therefore, the emergence time of the super-soft source can be used as a diagnostic for ejecta mass \citep{Henze11, Schwarz11}.  

Observations of novae in the $\sim$1--10 keV energy range also reveal harder X-ray emission on timescales of days to years after outburst \citep[e.g.,][]{Mukai08}. During the novae in RS~Oph and V407~Cyg, both of which have red giant secondaries, hard X-ray emission was detected at early times and attributed to the interaction of the nova ejecta with the dense wind of the companion \citep{Sokoloski06, Nelson12}.  In systems with less evolved donors (and hence lower density circumbinary environments), internal shocks within the ejecta have been proposed as the origin for hard X-ray emission \citep[see][and references therein]{Mukai08}. In these cases, hard X-ray emission can tell us about the structure of the ejecta as the nova outburst progresses \citep{O'Brien94}.  In the last three years, novae have been identified as a new class of GeV gamma-ray transient by {\it Fermi}/LAT, indicating that the shock interactions in novae are capable of accelerating particles to relativistic speeds (\citealt{Abdo10, Hill13}; see also \citealt{Tatischeff_Hernanz07}).  Observations of novae in the 1--10 keV range are required to characterize these shocks and fully understand the gamma-ray production mechanism. We note that T~Pyx was \emph{not} detected with \emph{Fermi} (C.~Cheung 2013, private communication), making it a useful comparison case for the study of why some novae produce detectable gamma-rays while others do not.

In this paper, we discuss both modes of X-ray emission during the 2011 outburst of T~Pyx. In Section 2 we discuss the X-ray observations and data reduction; \emph{Swift} monitoring reveals the X-ray evolution at high cadence, while our single epoch of \emph{Suzaku} spectroscopy provides high signal-to-noise on Day 194. In Section 3, we describe the \emph{Swift} X-ray light curve, and Section 4 presents our spectral analysis of the \emph{Swift} and \emph{Suzaku} data. Section 5 analyzes the observed hard X-ray component and concludes that it is likely produced by a shock within the ejecta, rather than interaction between the nova and pre-existing circumbinary material. Section 6 presents a super-soft component with a relatively cool temperature and late turn-on time (compared to other recurrent novae). In Section 7, we discuss how these X-ray results align with optical and radio observations, and suggest that all three wavelength regimes support a second, massive, delayed ejection in T~Pyx. We conclude in Section~8.

Throughout the paper, we take 2011 April 14 (MJD = 55665) as $t_{0}$ or Day 0, the beginning of optical rise and the start of the outburst \citep{Waagan11, Schaefer13}.  We also assume a distance to T~Pyx of $4.8\pm0.5$ kpc \citep{Sokoloski13}.

\section{Observations and Data Reduction} \label{data}

\subsection{\textit{Swift}}
The 2011 outburst of T~Pyx was monitored frequently with the {\it Swift} satellite as part of the ongoing monitoring of novae by the \emph{Swift} Nova CV Group \citep{Kuulkers11b, Kuulkers11a, Osborne11a}.  A series of 221 {\it Swift} X-ray Telescope (XRT) observations of T~Pyx were carried out between 2011 April 14 (Day 1) and 2012 April 17 (Day 369), resulting in one of the most detailed X-ray light curve of a nova obtained to date.  All observations were made in photon counting (PC) mode. The total XRT exposure time was $\sim$271 ks, with a median snapshot duration of $\sim$1 ks.  {\it Swift} observed T~Pyx daily during the first month of the outburst, and then reduced the cadence to 2--3 times per week for the subsequent three months.  Once the X-ray emission became bright at the start of September 2011 (see Section 3 below), the cadence was once more increased to daily, and remained so for the next two months.  The observing cadence was reduced at late times, ultimately resulting in weekly observations after January 2012, through April 2012 (Days 262 to 369).  

We created grade 0--12 event light curves and time-resolved hardness ratios using the {\it Swift}/XRT products generator developed by the {\it Swift} group at the University of Leicester\footnote{http://www.swift.ac.uk/user\_objects/}.  This web-based tool can create X-ray light curves, spectra and images of any object that has been observed with the XRT, and makes use of tools originally developed for automatic reduction of gamma-ray burst observations.  Details of these products are given in \citet{Evans09}.   We used an adaptive binning strategy to maximize the count rate and hardness ratio information in the light curves.  Prior to Day 124, and after Day 200, the data were binned to give a minimum of 20 counts per bin.  In between these days while the source was bright, the data were binned by observation (i.e. all snapshots with a single observation ID).  We also created spectra for certain time intervals during the outburst (see Section \ref{swiftspec}), again using the XRT products generator.  We obtained the appropriate response matrix file (RMF), in this case swxpc0to12s6$\_$20010101v013.rmf, from the calibration database. The resulting spectra were binned to have a minimum of one count per bin, and then modeled in XSpec using the Cash statistic, a maximum likelihood-based statistic for Poisson data \citep{Cash79}.

\subsection{\textit{Suzaku}}
\begin{figure*}[tbp]
\begin{center}
\includegraphics[height=4.0in]{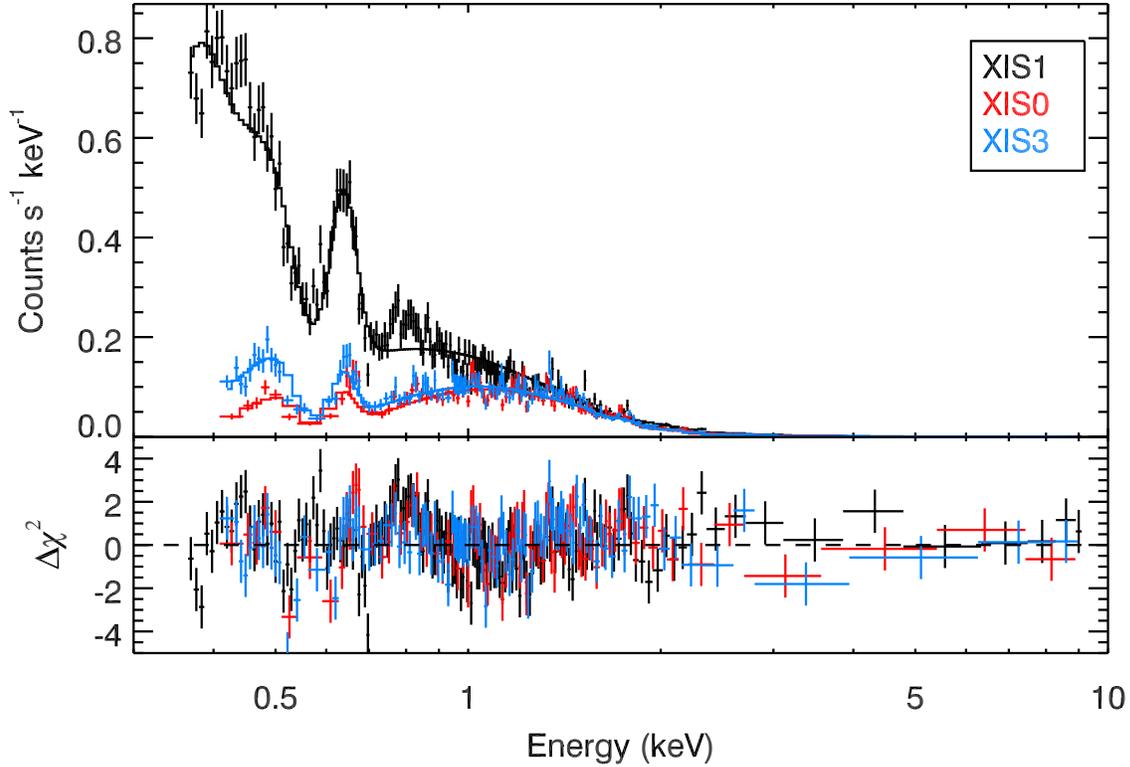}
\vspace{0.5cm}
\caption{The X-ray spectrum of T~Pyx as observed with {\it Suzaku} on Day 194 in three different XIS units (colored as black, red, and blue lines). The data are shown in the upper panel as points, overlaid with the best-fit model (the sum of a blackbody, a thermal bremsstrahlung plasma, and a Gaussian line) in solid lines. The parameters of this model can be found in Table \ref{tab:suzaku}. Residuals in the fit to the data are shown in units of sigma in the lower panel. The data shown here are without the energy offset discussed in Section 4.1, to illustrate the offset in energy calibration between XIS1 and XIS0/XIS3.
}
\label{figsuzaku}
\end{center}
\end{figure*}

Given the X-ray rise reported by \citet{Osborne11a}, we requested and were granted a Director's Discretionary Time observation of T~Pyx with the {\it Suzaku} observatory.  The exposure was carried out on 2011 October 25 (Day 194).  Here, we focus on the data obtained with the X-ray Imaging Spectrometer (XIS) in the 0.3--10 keV energy range.  All three functioning XIS units were operated in the full-window imaging mode, obtaining X-ray event data every 8 seconds over the full 19\arcmin\ by 19\arcmin\ field of view.  After standard screening\footnote{http://heasarc.gsfc.nasa.gov/docs/suzaku/processing/criteria\_xis.html?}, the net exposure time was $\sim$38,600 s for each XIS unit.  We extracted the source spectra from a 180\arcsec\ region centered on T~Pyx using XSelect v2.4b.  Background spectra were extracted from annular regions also centered on the source, with inner radius 240\arcsec\ and outer radius 420\arcsec\ for both XIS1 and XIS3, while the outer radius was 390\arcsec\ for XIS0 (the smaller size is to avoid dead regions of the chip) .  We created response files using the {\tt xisrmfgen} and {\tt xisarfgen} ftools. The most recent version of the contamination model (version 20120719) was used to calculate the effective areas.  The resulting spectra were binned to have a minimum of 30 counts per bin in order to facilitate the use of $\chi^{2}$ statistics in determining the best fit spectral model (Figure \ref{figsuzaku}).

\section{X-ray Light Curve of the 2011 Outburst of T~Pyx}

We present the {\it Swift} XRT light curve in the middle panel of Figure \ref{rxo}, as well as the V-band AAVSO optical light curve (upper panel) and a three-frequency radio light curve obtained with the Karl G.~Jansky Very Large Array (VLA) in the lower panel. The \emph{Swift} XRT light is also plotted in Figure \ref{hratio}a with a linear time axis.

Individual X-ray observations during the first 100 days of the outburst mainly resulted in non-detections, with the notable exception of the detections between Days 14--20. T~Pyx is detected as a faint hard X-ray source in this third week after outburst (Figure \ref{hratio}b), but then the X-rays fade out of detectability again for the next three months.

On Day 117, T~Pyx was seen at 0.003 cts~s$^{-1}$, and by Day 125 it had increased in count rate by a factor of ten. T~Pyx then entered a sun-angle constraint, and was not observed again until 2011 September 3 (Day 142).  This first visit back to T~Pyx revealed that the source had increased in count rate by another order of magnitude, to around 0.4 cts~s$^{-1}$. The 0.3--10 keV count rate remained high, although variable, until Day $\sim$180, at which point it began to fade systematically over the following 200 days.  By the time of the last monitoring observation with \emph{Swift} on 2012 April 17, the source count rate was $\sim$0.07 cts~s$^{-1}$.

\section{X-ray Spectra of the 2011 Outburst of T~Pyx}

\subsection{A Deep View on Day 194 with {\it Suzaku}} \label{suzaku}
Although the {\it Swift} observations were carried out with high cadence, in general the continuous observation durations were rather short and resulted in limited counting statistics for spectroscopy.  The {\it Suzaku} observation on Day 194 resulted in $\sim$22,000 counts between all three XIS units (Figure \ref{figsuzaku}).  We use this deep observation to characterize the X-ray emission in as much detail as possible, and to inform our model for the lower count statistic {\it Swift} data.  

We modeled the spectra extracted from all three XIS units jointly in Xspec version 12.8.0m \citep{Arnaud96}, and present the resulting parameters in Table \ref{tab:suzaku}.  In all models, a constant was included in order to allow for uncertainty in the calibration of the relative effective areas of the three  CCDs (typically of order 5-10\%).  

To account for absorption by the ISM we use the {\tt tbnew} model of \citet{Wilms00}, and assume the cross-sections of \citet{Verner96} and the interstellar medium (ISM) abundances of \citet{Wilms00}.  A range of values for the degree of interstellar absorption towards T~Pyx have been given in the literature.  \citet{Gilmozzi_Selvelli07} find a value of $E(B-V) = 0.25 \pm 0.02$ from {\it IUE} ultraviolet spectra of T~Pyx obtained during quiescence, which corresponds to a value of $N(H) \approx  (1.3\pm0.1) \times 10^{21}$~cm$^{-2}$ (assuming the correlation between $N(H)$ and $E(B-V)$  of \citealt{Predehl_Schmitt95}). \citet{Shore11} used diffuse interstellar bands observed during the 2011 outburst to determine $E(B-V) = 0.49\pm0.17$ mag for T~Pyx, implying $N(H) \approx (2.6\pm0.9) \times 10^{21}$~cm$^{-2}$.  Finally, \citet{Godon14} determined an intermediate reddening value from UV spectroscopy, $E(B-V) = 0.35$ mag.

\begin{deluxetable*}{lcccc}[t]
\tabletypesize{\footnotesize}
\tablecaption{ \label{tab:suzaku}
Best-Fit Model Parameters for Day 194 \emph{Suzaku} Spectrum\tablenotemark{1}}
\tablehead{ & {\tt tbnew*(bb+brems+gauss)} & {\tt tbnew*(bb+vapec)$_a$} & {\tt tbnew*(bb+vapec)$_b$} & {\tt tbnew*(bb+vnei)}}
\startdata
N(H) (10$^{21}$ cm$^{-2}$) &  0.27 $\pm$ 0.19 & 2.0 $\pm$ 0.02 & 0.61 $\pm$ 0.02 & 0.5 $\pm$ 0.02 \\
\\
 kT$_{\rm BB}$ (eV) & 47 $\pm$ 2 & 35 $\pm$ 2 & 44 $\pm$ 2 & 46 $\pm$ 2 \\
 norm$_{\rm BB}\tablenotemark{2}$ & 0.0016$^{+0.0010}_{-0.0006}$ & 0.083$^{+0.002}_{-0.001}$ & 0.0032 $\pm$ 0.0020 & 0.0026$^{+0.0018}_{-0.0007}$ \\
 \\ 
kT$_{\rm br}$ (keV) &  0.72 $\pm$ 0.03 & \nodata & \nodata & \nodata \\
norm$_{\rm br}$\tablenotemark{3} (10$^{-3}$) & 2.5 $\pm$ 0.2  & \nodata & \nodata & \nodata \\
\\
kT$_{\rm vapec}$ (keV) & \nodata & 0.70 $\pm$ 0.02 & 0.73 $\pm$ 0.04 & \nodata \\
norm$_{\rm vapec}\tablenotemark{4}$ (10$^{-3}$) & \nodata & 0.25 $\pm$ 0.03 & 7.4 $\pm$ 0.5 & \nodata\\
Z/Z$_{\odot}$ & \nodata & 1.0                          & $<$0.003 & \nodata \\
O/O$_{\odot}$ & \nodata & 150$^{+6}_{-4}$ & 1.6$^{+0.3}_{-0.2}$ & \nodata \\
\\
kT$_{\rm vnei}$ (keV) & \nodata & \nodata & \nodata & 0.70 $\pm$ 0.02\\
norm$_{vnei}\tablenotemark{4}$ (10$^{-3}$) & \nodata & \nodata & \nodata  & 7.4$^{+1.0}_{-1.2}$  \\
Z/Z$_{\odot}$ & \nodata & \nodata                         & \nodata & 0.004$^{+0.002}_{-0.003}$ \\
O/O$_{\odot}$ & \nodata & \nodata & \nodata & 0.8$^{+0.4}_{-0.3}$ \\
$\tau$ (10$^{11}$ s cm$^{-3}$) & \nodata & \nodata & \nodata & 2.9$^{+0.8}_{-0.4}$ \\
\\
E$_{\rm gauss}$ (keV) & 0.643$^{+0.002}_{-0.003}$ & \nodata & \nodata & \nodata \\
norm$_{\rm gauss}$\tablenotemark{5} (10$^{-3}$) & 0.60$^{+0.09}_{-0.08}$ & \nodata & \nodata & \nodata \\
\\
XIS 1 normalization & 1.0 & 1.0 & 1.0 & 1.0 \\
XIS 0 normalization & 0.92 & 0.92 $\pm$ 0.03 & 0.92 $\pm$ 0.03 & 0.92 $\pm$ 0.03 \\
XIS 3 normalization & 0.87 & 0.85 $\pm$ 0.03 & 0.86 $\pm$ 0.03 & 0.86 $\pm$ 0.02 \\
\\
$\chi^{2}$ & 657.07 & 895.58 & 665.24 & 653.34\\
D.O.F & 539 & 540 & 539 & 537 \\
\enddata
\tablenotetext{1}{All quoted uncertainties are 90\% confidence intervals.}
\tablenotetext{2}{norm$_{BB}$ = $\frac{L_{39}}{D_{10}^{2}}$, where $L_{39}$ is the bolometric luminosity of the source in units of 10$^{39}$ erg s$^{-1}$, and $D_{10}$ is the distance to the source in units of 10 kpc.}
\tablenotetext{3}{norm$_{br}$ = $\frac{3.02 \times 10^{-15}}{4\pi D^{2}} \int n_{e} n_{i} dV$, where $D$ is the distance to the source in units of cm and $n_{e}$ and $n_{i}$ are the electron and ion number densities, respectively, in units of cm$^{-3}$.}
\tablenotetext{4}{norm$_{vapec}$ = $\frac{10^{-14}}{4\pi D^{2}} \int n_{e} n_{i} dV$, where the parameters have the same meaning as in the bremsstrahlung model. }
\tablenotetext{5}{norm$_{\rm gauss}$ = Total photons cm$^{-2}$ s$^{-1}$ in the gaussian line.}
\end{deluxetable*} 

The {\it Suzaku} spectrum is clearly complex.  A distinct soft component is observed below 0.5 keV and is most obvious in the XIS1 data, which has the highest sensitivity at low energies. A second, harder continuum component extends out to $\sim$6 keV.  Finally, there is a rather striking resolved emission line at $E \approx 0.65$ keV that we identify as \ion{O}{8} Ly$\alpha$.  There are no obvious emission lines at any other location in the spectrum.   

We begin our exploration of the spectrum with a simple model that accounts for the main features seen in the data: the sum of a blackbody, a thermal plasma, and a Gaussian emission line (model {\tt tbnew*(bb+brems+gauss)} in Table \ref{tab:suzaku}). This model is not strictly physical, as the soft component is expected to show features typical of a white dwarf atmosphere and the plasma is expected to cool primarily through lines, but it does provide a useful starting point given the limited energy resolution of the spectrum.   This simple model provides a reasonable fit to the data, with $\chi^{2}/\nu$ = 1.26 ($\nu$ = 539).  The Gaussian line is clearly required to account for the feature at $\sim$0.65 keV.  However, the best-fit energy of the emission line is found to be $0.643^{+0.002}_{-0.003}$ keV, implying that the \ion{O}{8} line is redshifted by $\sim$5,000 km s$^{-1}$.  This velocity is at odds with the results of \citet{Tofflemire13}, who reported that the \ion{O}{8} Ly$\alpha$ line is blue-shifted by $\sim -400$ km s$^{-1}$ in a {\it Chandra} Low Energy Grating Spectrograph (LETG) observation taken a few months later (Day 210). The observed redshift is also larger than velocities derived from optical spectroscopy at any point during the outburst.

A careful examination of the residuals around the emission line in all three XIS units reveals a shift to lower energies in the XIS1 data, possibly indicating a problem with the energy calibration of the observation.   At low energies ($<$ 1 keV), uncertainties in the energy scale calibration come primarily from the determination of the CCD zero-level, and it is possible that this was more uncertain for the XIS1 CCD than the others. We investigated the presence of this energy offset in XIS1 relative to the other two CCDs using the {\tt gain fit} command in XSpec.  Fitting all three spectra independently with an absorbed blackbody model, we find evidence of an offset in the energy scale of XIS1 by approximately $-8$ eV relative to XIS0 and XIS3.  Including this additional energy offset in the model fit decreases both the reduced $\chi^2$ value of the fit, and the residuals around the emission line in the XIS1 data.  We therefore include an energy offset of $-8$ eV for the XIS1 spectrum in all of our fits moving forward.  This energy offset is consistent with the estimated uncertainty in the energy scale below 1 keV of $<$10 eV\footnote{\url{http://web.mit.edu/iachec/meetings/2012/Presentations/Miller.pdf}}, although it may indicate that the error in the energy scale determination for this observation was larger than typical values. Comparing fits with and without the offset, we find very little difference in the values associated with the blackbody and bremsstrahlung components, but the energy of the line shifts to higher energies with the inclusion of the energy offset.  We therefore caution that the uncertainties in the line energy are larger than the statistical errors from the fit alone, likely of order 10 eV (or 4,500 km s$^{-1}$ in velocity space).  

The best-fit parameters obtained for the simple blackbody+thermal plasma+Gaussian line model, obtained with the inclusion of the additional XIS1 energy offset, are shown in Table \ref{tab:suzaku}.  This model is presented in Figure \ref{figsuzaku}, overplotted on the observed spectra.  The best-fit N(H) value for this model is only $(2.7 \pm 1.9) \times 10^{20}$ cm$^{-2}$,  much lower than the value expected from optical and UV estimates of reddening.   The blackbody temperature, kT$_{\rm BB}$, is found to be 47 $\pm$ 2 eV (545,400 $\pm$ 23,000 K, 90\% confidence level), with the main constraint coming from the well-sampled Wien tail between 0.4 and 0.6 keV.  The normalization of this component implies a blackbody luminosity of $(3.7^{+3.5}_{-1.8}) \times 10^{35}$ erg s$^{-1}$.  The plasma temperature is found to be 0.72 $\pm$ 0.03 keV, or $(8.4 \pm 0.4) \times 10^{6}$ K.  The normalization of this component corresponds to a volume emission measure of  $EM_V = n^{2}\, V$ = (2.2$^{+0.2}_{-0.1}) \times 10^{57}$ cm$^{-3}$, or an X-ray luminosity of (8.9$^{+1.0}_{-0.6}$) $\times$ 10$^{33}$ erg s$^{-1}$.

\begin{deluxetable*}{ccccccccccc}[h]
\tabletypesize{\footnotesize}
\tablecaption{ \label{tab:swift}
Best-Fit Model Parameters for Weekly-Averaged \emph{Swift} X-ray Spectra\tablenotemark{1}}
\tablehead{Date Range & Time since & N(H) & kT$_{\rm BB}$ & norm$_{\rm BB}$\tablenotemark{2} & kT$_{\rm br}$ & norm$_{\rm br}$\tablenotemark{2} & E$_{\rm gauss}$ & norm$_{\rm gauss}$\tablenotemark{2} & c-stat & D.O.F. \\ 
(UT) & Outburst (Days) & (10$^{21}$ cm$^{-2}$) & (eV) & & (keV) & (10$^{-3}$) & (keV) & (10$^{-3}$) & & }
\startdata
2011 Sep 03--09       & 142--149 & 2.6$^{+0.5}_{-0.5}$ & 32$^{+4}_{-3}$     & 0.61$^{+1.76}_{-0.46}$             & 1.20$^{+0.14}_{-0.13}$ & 3.4$^{+0.6}_{-0.5}$ & 0.666$^{+0.013}_{-0.014}$ & 1.39$^{+0.59}_{-0.33}$ & 268.1 &  267 \\
\\
2011 Sep 10--16       & 149-156 & 2.2$^{+0.6}_{-0.6}$ & 40$^{+6}_{-5}$     & 0.053$^{+0.257}_{-0.040}$       & 1.21$^{+0.20}_{-0.17}$ & 3.1$^{+0.7}_{-0.6}$ & 0.666\tablenotemark{3} & 0.94$^{+0.86}_{-0.60}$ & 220.8 &  222 \\
\\
2011 Sep 17--23       & 156--163 & 1.5$^{+0.5}_{-0.4}$ & 45$^{+6}_{-5}$     & 0.013$^{+0.027}_{-0.008}$       & 1.17$^{+0.15}_{-0.16}$ & 3.3$^{+0.6}_{-0.5}$ & 0.666\tablenotemark{3} & 0.53$^{+0.50}_{-0.38}$ & 273.7 &  243 \\
\\
2011 Sep 24--30       & 163--170 & 1.8$^{+0.5}_{-0.5}$ & 43$^{+6}_{-5}$     & 0.019$^{+0.013}_{-0.13}$             & 1.09$^{+0.16}_{-0.15}$ & 3.4$^{+0.8}_{-1.6}$ & 0.666\tablenotemark{3} & 0.79$^{+0.64}_{-0.46}$ & 237.7 &  232 \\
\\
2011 Oct 01--07        & 170--177 & 2.1$^{+0.7}_{-0.5}$ & 39$^{+6}_{-5}$     & 0.061$^{+0.312}_{-0.047}$       & 1.01$^{+0.25}_{-0.07}$ & 3.1$^{+0.9}_{-0.6}$ & 0.666\tablenotemark{3} & 1.50$^{+0.86}_{-0.71}$ & 201.78 &  215 \\
\\
2011 Oct 08--14        & 177--184 & 1.4$^{+0.4}_{-0.4}$ & 48$^{+6}_{-6}$     & 0.0071$^{+0.018}_{-0.004}$       & 1.00$^{+0.15}_{-0.14}$ & 23.0$^{+0.7}_{-0.6}$ & 0.666\tablenotemark{3} & 0.46$^{+0.46}_{-0.35}$ & 229.4 &  216 \\
\\
2011 Oct 16--21        & 184--191 & 0.7$^{+0.4}_{-0.3}$ & 58$^{+7}_{-8}$     & 0.0010$^{+0.0019}_{-0.0005}$ & 0.92$^{+0.19}_{-0.14}$ & 2.0$^{+0.6}_{-0.5}$ & 0.666\tablenotemark{3} & $<$0.82 & 161.0 & 189 \\
\\
2011 Oct 23--29        & 191--197 & 1.2$^{+0.9}_{-0.6}$ & 53$^{+13}_{-13}$   & 0.0022$^{+0.021}_{-0.0017}$       & 0.79$^{+0.27}_{-0.17}$ & 2.3$^{+1.5}_{-0.9}$ & 0.666\tablenotemark{3} & 0.40$^{+0.64}_{-0.38}$ & 129.8 & 146 \\
\\
2011 Oct 30-- & 197--204 & 1.2$^{+0.9}_{-0.7}$ & 53$^{+16}_{-13}$ & 0.0010$^{+0.006}_{-0.0008}$     & 0.58$^{+0.16}_{-0.10}$ & 2.6$^{+1.8}_{-1.1}$ & 0.666\tablenotemark{3} & 0.39$^{+0.47}_{-0.26}$ & 122.7 & 148 \\
Nov 05 
\enddata
\tablenotetext{1}{All quoted uncertainties are 90\% confidence intervals.}
\tablenotetext{2}{norm$_{BB}$, norm$_{br}$, and norm$_{gauss}$ as defined in Table \ref{tab:suzaku}. }
\tablenotetext{3}{Parameter was fixed in model fit.}
\end{deluxetable*}

We have accounted for the presence of the \ion{O}{8} Ly$\alpha$ line in this simple model with an additional Gaussian.  However it is more realistic to attempt to fit the data with a model that produces both continuum and line emission.  Therefore, we also modeled the harder component with the APEC models of collisional ionization equilibrium plasmas (\citealt{Foster12}).  A solar abundance model gives a very poor fit to the data---the lines of He-like Si and S, as well as the Fe L-shell emission around 0.9 keV are much too strong in the model compared to the data, while the \ion{O}{8} Ly$\alpha$ line is too weak.  If we allow the abundances of all metals to vary in tandem (while keeping their relative abundances at solar values), the model can achieve a good fit to the data with no residuals at the Si and S lines if the best fit metallicity is extremely subsolar ($Z \approx 0.01\, Z_{\odot}$).  However, such a model still fails to reproduce the \ion{O}{8} Ly$\alpha$ line observed in the data, and results in abundances seemingly at odds with the report of approximately solar abundances in the larger scale T~Pyx ejecta \citep{Williams82}.

We obtain a better, but still inadequate, fit to the data if the abundance of O in the plasma is extremely super-solar, while assuming solar abundance for all other elements (model {\tt tbnew*(bb+vapec)$_a$} in Table \ref{tab:suzaku}).  Leaving both the O abundance of the plasma to vary freely and keeping all other abundances fixed at solar values relative to hydrogen, we find  $\chi^{2}/\nu = 1.7$ ($\nu = 540$) for an oxygen abundance of $150^{+6}_{-4}$ times solar.  The temperature of this plasma is similar to that found for the simple bremsstrahlung model (kT$_{\rm br} = 0.70 \pm 0.02$  keV, or $8.1 \times 10^{6}$ K).  The temperature of the blackbody is slightly lower (kT$_{\rm BB}$ = 35 eV, T$_{\rm BB}$ = 406,000 K). The most marked difference is in the N(H), which is found to be more in line with the optical and UV values, N(H) = $(2.0 \pm 0.02) \times 10^{21}$ cm$^{-2}$.  

The large oxygen abundance inferred by the strong emission line in our spectrum could indicate that the shocked material has been significantly polluted by material from the white dwarf. Interestingly, \citet{Tofflemire13} find evidence for a strong overabundance (relative to solar) of N in their high spectral resolution \emph{Chandra}/LETG data from Days 210 and 235; our earlier \emph{Suzaku} data unfortunately do not have sufficient resolution at low energies to explore this finding. Both oxygen and nitrogen over-abundance are suggestive of the presence of dredged-up white dwarf material in the thermal plasma. 

We also tested an alternative possibility; that the X-rays come from a primarily oxygen-rich plasma that has very little metal content.  We fit an absorbed blackbody + vapec model to the data, but this time allowed all elemental abundances to vary freely (model {\tt tbnew*(bb+vapec)$_b$} in Table \ref{tab:suzaku}).  Given the lack of obvious spectral features, we kept the relative abundances of He, C, N and all metals with higher atomic numbers than O fixed at their solar ratios relative to Fe, and allowed the abundance of Fe to vary freely.  The abundance of oxygen was allowed to vary independently.  This model results in a significantly better fit to the data than the previous model, with $\chi^{2}/\nu = 1.2$ ($\nu = 539$).  The abundances of all metals other than O are essentially zero, and the abundance of oxygen relative to hydrogen is $\sim$60\% enhanced relative to solar values.  The temperatures of the two components are similar to those found for other models. 

Alternatively, the bright oxygen line could be an indication that the plasma is underionized and has not yet come into collisional equilibrium (i.e. in non-equilibirum ionization; NEI). To test this scenario, we use the {\tt vnei} model in XSpec.  The overall spectrum is modeled as the sum of an absorbed blackbody plus NEI plasma (Model {\tt tbnew*(bb+vnei)} in Table \ref{tab:suzaku}).  Keeping all elemental abundances at their solar values, we find no fits with $\chi^{2}/\nu < 1.7$ ($\nu$ = 538).  Allowing the oxygen abundance to vary freely, but keeping all other elements fixed at solar does not improve the fit much ($\chi^{2}/\nu < 1.6$, $\nu$ = 537).  Allowing all other elements to vary in tandem from their solar values, as we did for the {\tt vapec} model above, does result in a better fit ($\chi^{2}/\nu =1.2$, $\nu$ = 537), and results in a similar ratio of oxygen to other metals (see Table \ref{tab:suzaku} for model parameters).  The best-fit ionization age, 2.9 $\times$ 10$^{11}$ s cm$^{-3}$, is close to the equilibrium value for a plasma with kT = 0.7 keV \citep{Smith10}, suggesting that non-equilibrium ionization physics is not the primary reason for the strong oxygen relative to other metals.  

In summary, the key results of our modeling of the Suzaku spectrum are (1) all models require both a soft and hard component to describe the emission, and (2) only a plasma that is significantly enhanced in oxygen relative to all other elements can reproduce the observed {\it Suzaku} spectrum at E $>$ 0.5 keV.

\begin{figure*}[t]
\begin{center}
\includegraphics[height=3.5in]{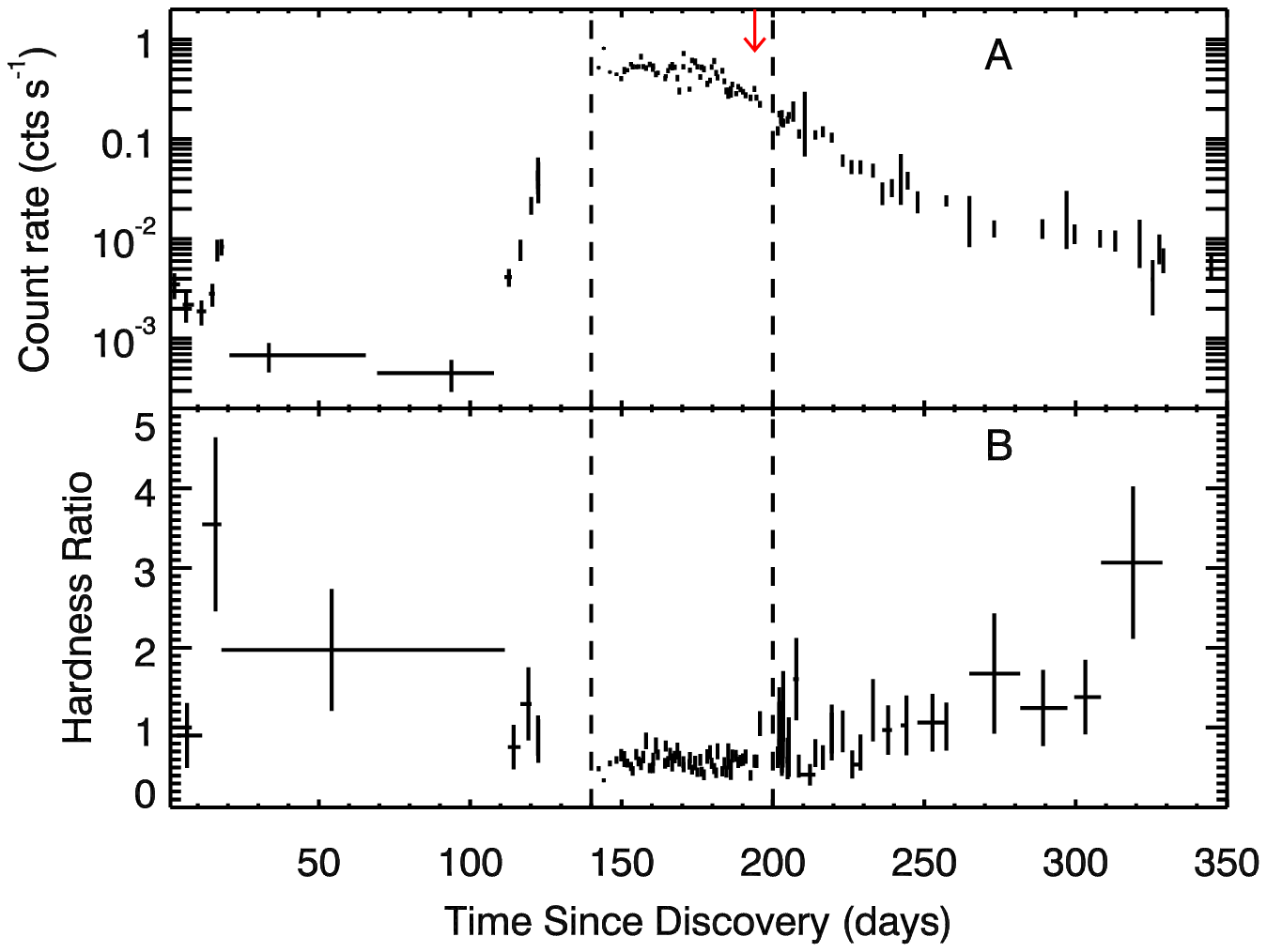}
\includegraphics[height=5in]{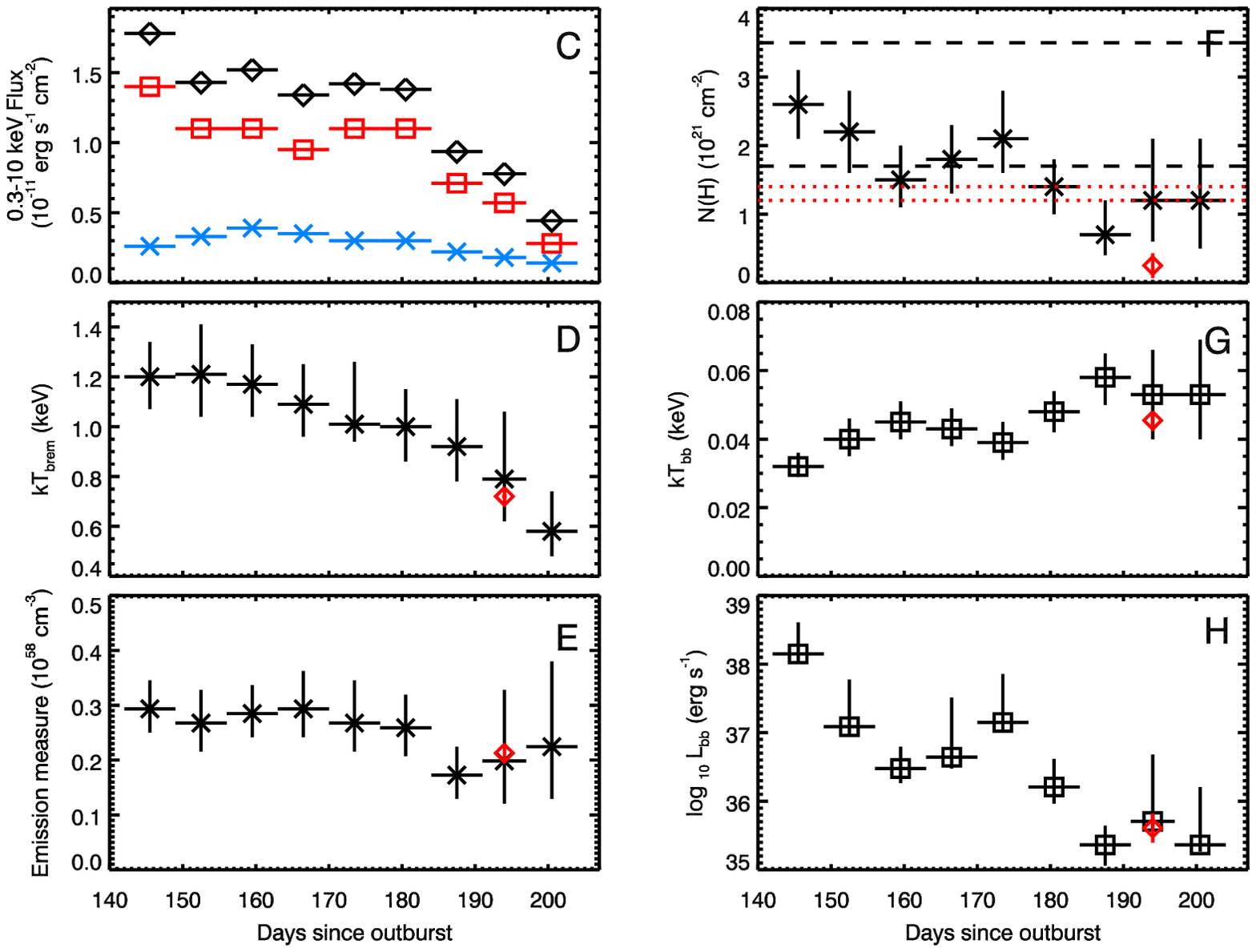}
\caption{{\it a:} \emph{Swift}/XRT light curve of T~Pyx in the 0.3--10 keV energy range (as in Figure 1, this time with a linear time axis). The red arrow marks the time of the {\it Suzaku} observation, and the dashed vertical lines mark the period for which we analyze the {\it Swift} spectra. {\it b:} X-ray hardness ratio (defined as the ratio of the 0.55--10 keV to 0.3--0.55 keV count rate) as a function of time. {\it c:} X-ray flux in 0.3--10 keV range derived from weekly averaged {\it Swift} spectra. Black diamonds show the total flux, red squares mark the blackbody component, and blue crosses denote the bremsstrahlung flux. {\it d} and {\it e:} Temperature and volume emission measure evolution of the thermal plasma component. {\it f:} The evolution of the absorbing column density.  The range of interstellar absorbing column derived by \citet{Shore11} is shown as black dashed lines, and the \citet{Gilmozzi_Selvelli07} measurement is bracketed by red dotted lines.  {\it g} and {\it h:} The evolution of the blackbody temperature and luminosity. The best fit values from the {\it Suzaku} spectrum are shown as red diamonds in panels {\it d--h}.}
\label{hratio}
\end{center}
\end{figure*}

\subsection{Spectral Fits from {\it Swift}} \label{swiftspec}
To examine the time evolution of the distinct components we identified in the {\it Suzaku} spectrum, we extracted and modeled weekly-averaged spectra from the {\it Swift} observations.  The flux level and total number of counts before Day 142 are too low to allow useful spectra to be produced.  Likewise, we find that after Day 206 the weekly spectra have $<$100 total counts, and so no stable model fits are found. We therefore restrict our spectroscopic analysis of the {\it Swift} data to weekly averaged spectra between Day 142 and Day 206 (2011 Sep~3--Nov~6).  At each epoch we model the fit with the simple empirical {\tt tbnew*(bb+brems+gauss)} model developed for the {\it Suzaku} data, i.e., the sum of a blackbody, a Gaussian emission line, and thermal bremsstrahlung absorbed by the interstellar medium. Visual inspection of the the first few weekly spectra shows an enhancement of counts at the position of the \ion{O}{8} line found in the \emph{Suzaku} spectrum, justifying the inclusion of the Gaussian line component. We fixed the energy of the Gaussian line because of the lower number of counts in the \emph{Swift} spectra, and allowed the line flux to vary freely.  Given the uncertainties in the {\it Suzaku} energy calibration discussed in Section 4.1, we fixed the energy of the line to the best-fit line energy found from the \emph{Swift} Sept 3--9 spectrum (0.666 keV).   Initial values for all other parameters are taken from the {\it Suzaku} best-fit model and then allowed to vary freely.  

The resulting best fit model parameter values and their 90\% confidence interval uncertainties are shown in Figure \ref{hratio} and listed in Table \ref{tab:swift}. Where the \emph{Swift} and \emph{Suzaku} observations overlap, \emph{Swift}-derived parameters are consistent with measurements from the the deeper {\it Suzaku} observation (\emph{Suzaku} parameters from the {\tt tbnew*(bb+brems+gauss)} model shown as red diamonds in Figure \ref{hratio}d--h).  The only exception is the N(H) value, which is considerably lower for the {\it Suzaku} best-fit model than the overlapping {\it Swift} data.  However, we note that the error bars shown on this plot are the statistical uncertainties from the model fit only---they do not include a larger systematic uncertainty related to the choice of model (see Table 1 for range of N(H) values found). 

Over the $\sim$60 days where T~Pyx is X-ray bright, we see evolution in the spectrum and flux that is dominated by changes in the super-soft blackbody component (Figure \ref{hratio}c). The temperature of the thermal plasma component is roughly constant between Days 140 and 180, and then declines by a factor of $\sim$2 by Day 206 (Figure \ref{hratio}d).  The volume emission measure over the same time period does not vary significantly (Figure \ref{hratio}e), and so the decline in temperature results in a modest drop in flux from the hard X-ray component (see blue points in Figure \ref{hratio}c).   

The drop in the total flux by a factor of $\sim$2 between Day 142--206 is mainly attributable to the blackbody component (Figure \ref{hratio}c). The decline in the super-soft light curve can also be seen in Figure \ref{hratio}h, where we see that the blackbody luminosity is lower after Day 175 than before. This decline in flux is despite a roughly constant value of the absorbing column density (Figure \ref{hratio}f; consistent with the interstellar values estimated by \citealt{Gilmozzi_Selvelli07} and \citealt{Shore11}) and a blackbody temperature evolution which shows no signature of decline (Figure \ref{hratio}g). 

In all but one of the spectra, the flux of the emission line component is non-zero (Table 2). The flux appears to drop slightly in the first three weekly-averaged spectra, and then re-brightens to its highest level in the Oct~1--7 spectrum.  The line then fades again; we only obtain an upper limit to the line flux in the Oct~16--21 spectrum. The line flux is non-zero in the last two epochs, and $\sim$60\% smaller compared to September measurements.

We stress that the X-ray properties shown in Figure \ref{hratio}c--h are only obtained during the plateau in the X-ray light curve, Days 142--206. The count rates measured before Day 117 and after Day $\sim$250 are orders of magnitude lower, and so we can not glean spectral information at this time due to the poor signal-to-noise in these observations.  However, we can infer something about the spectral properties from the hardness ratio plot (Figure 3b).  We choose 0.55 keV as the break energy between the soft and hard bands, since our {\it Suzaku} spectral fits imply that all flux above this energy originates in the thermal plasma component.  At very early times, the spectrum appears to be quite hard, particularly at the time of the first X-ray detections around Days 14--20.  The spectrum then becomes softer as the source gets brighter.  Although the error bars are quite large, the hardness ratio during the X-ray rise is higher than at peak brightness, which suggests that the hard X-ray component rose before the super-soft emission emerged.  At late times, we observe a slight trend towards higher hardness ratios, presumably occurring as the super-soft continuum fades. 

\section{The Origin of Hard X-ray Emission in T~Pyx}\label{hardx}

The hard X-rays ($E >$ 1 keV), modeled in the previous section as thermal plasma and seen in all \emph{Swift}/XRT detections, indicate the presence of a long-lived shock. Hard X-ray emission in novae can originate either in external material that is swept up and shocked by the nova ejecta \citep{Sokoloski06,Nelson08,Nelson12}, or in shocks within the ejecta themselves \citep{O'Brien94, Mukai01}. In this section, we consider the possibility that the hard X-ray emission is produced by interaction between the nova ejecta and pre-existing circumbinary material. We will show that the timing of the hard X-ray component rules out plausible models for circumbinary material around T~Pyx as a significant source of X-rays. We therefore propose that an interaction \emph{within} the 2011 ejecta, between two distinct episodes of ejection, explains the hard X-rays.  

The temperature associated with the thermal plasma emission encodes information about the kinematics of the shock.  During the first week T~Pyx is visible after solar conjunction (Days 142--148, the first \emph{Swift} observations with good signal-to-noise; Table \ref{tab:swift}), the best-fit temperature of the hard X-ray emitting thermal plasma is T$_{\rm br} \approx 1.4 \times10^{7}$~K.  This temperature is comparable to those measured in hard X-ray components of other novae \citep{Lloyd92, Balman98, Mukai01}, and can be translated into a rough estimate of the shock velocity using the Rankine-Hugoniot jump conditions for a strong shock:
\begin{eqnarray}
v_s & =  &\sqrt{{16\ {\rm k T}_{\rm br}} \over {3\ \mu\, m_{H}}}\\
{{v_s}\over{1000~{\rm km~s}^{-1}}}  & = & \sqrt{{{\rm T}_{\rm br}} \over{1.4 \times 10^{7}~{\rm K}}}
\end{eqnarray} 
Here, $\mu$ is the mean molecular weight of the gas and $m_{H}$ is the mass of a hydrogen atom. We therefore estimate $v_s \approx 1,000$ km s$^{-1}$ for the X-ray emitting shock in T~Pyx. 

The maximum velocity of the ejecta is a critical parameter for our models of shocked X-ray emission in T~Pyx. We proceed with our discussion using H$\beta$ measurements of the maximum blueshift from Paper I; these imply a maximum expansion velocity of $\sim$1,900 km s$^{-1}$ on Day 2 and $\sim$3,000 km s$^{-1}$ on Day 69 (see also \citealt{Surina14}).

\subsection{Could Interaction with Circumbinary Material Produce the Hard X-Ray Component?}\label{cbm}

A rich circumbinary medium is clearly present around T~Pyx, in the form of a spatially-resolved, clumpy H$\alpha$+[\ion{N}{2}] remnant \citep{Williams82, Shara_etal97, Schaefer10a}. \citet{Contini97} propose that X-ray emission may be produced in T~Pyx when ejecta from nova outbursts interact with this nebula. \citet{Balman10} and \citet{Balman12} claimed a detection of spatial extension in X-ray images of T~Pyx in quiescence, as might be expected for such interaction. Subsequently however, this claim was called into question using re-analyzed high-resolution \emph{Chandra} X-ray images \citep{Montez12}.
 
 Interaction between the 2011 ejecta and the spatially resolved nebula is simple to rule out with timescale arguments. The H$\alpha$+[\ion{N}{2}] flux from the nebula peaks at 4.4$^{\prime\prime}$, or 0.1 pc \citep{Shara_etal97}. All nine of the nebula's ``shells" proposed by \citet{Shara_etal97} have radii $>1^{\prime\prime}$, or $6 \times 10^{16}$ cm. Similarly large radii, $2 \times 10^{17}$ cm, are expected for the ejecta from the 1966 outburst, if they expand at 1,900 km s$^{-1}$. Assuming the 2011 ejecta are traveling at 1,900--3,000 km s$^{-1}$, an encounter on Day 117 would imply that material from the 1966 ejection is traveling at just 20 km s$^{-1}$. In the unlikely case that there are ejecta from 1966 traveling at such a low velocity, we would expect the shell of 1966 ejecta to be very thick (spanning velocities 20--3,000 km s$^{-1}$), and the shock interaction to be prolonged over years. Instead, the rise and plateau of the hard X-ray light curve is contained to a relatively short period, Days $\sim$117--206 (Figure \ref{hratio}). Reasonable assumptions for the velocity and distribution of the 1966 ejecta therefore imply that they are much too distant to be encountered by the 2011 ejecta in the first year. 

 In addition to the ejecta from previous novae, there might be circumbinary material present at smaller radii, perhaps lost through a wind during the quiescent period, 1967--2011. Indeed, T~Pyx seems to have an unusually high accretion rate for a cataclysmic variable, so it is not unreasonable to expect such mass loss \citep{Selvelli08, Schaefer13, Godon14}. However, from UV spectroscopy during quiescence, there is no evidence of a strong or fast outflow from T~Pyx \citep{Gilmozzi_Selvelli07}. 

Given the low inclination of T~Pyx \citep{Uthas10}, perhaps the easiest way to ``hide" significant mass loss from the binary during quiescence is to have it leak out of an outer Lagrangian point and remain concentrated in the orbital plane; we might then expect mass loss to proceed at relatively modest speeds on order of the orbital velocity, $\sim$20 km s$^{-1}$ \citep{Uthas10}. The density of this equatorial circumbinary material should be highest near the binary, and fall off linearly with distance. This configuration of circumbinary material is in direct conflict with the primary late rise of the X-ray light curve. 
While the faint detections between Days 14--20 might be consistent with such an interaction, the X-ray non-detections between Days 20--117 imply a cavity surrounding T~Pyx, extending between radii $\sim4\times10^{14}-2 \times 10^{15}$ cm.
No simple picture of circumbinary material in T~Pyx explains a cavity at these radii, surrounded by a dense shell.
We therefore move on to consider an alternative explanation for the hard X-ray component in T~Pyx that rises four months after outburst: shocks within the 2011 nova ejecta.

\subsection{Hard X-rays Produced By Shocks Within the Ejecta} \label{intsh}

Shocks may be produced within the nova ejecta themselves, if the speed of ejection increases with time after the thermonuclear runaway \citep{O'Brien94}. In this case, the shock velocity ($v_s$) is equal to $4/3$ the differential velocity between the inner and outer ejecta components. In fact, we see increasing outflow velocities in T~Pyx over the first two months after outburst, as measured from optical spectroscopy (Paper I, \citealt{Surina14}). The difference between the velocity of material ejection on Day 2 and Day 69, as measured from the H$\beta$ profile ([3,000 -- 1,900] km~s$^{-1}$ = 1,100 km~s$^{-1}$), is consistent with the interaction velocity estimated from the X-ray temperature ($3/4 v_s \approx 800$ km~s$^{-1}$).

Let us simplify the complex picture painted by optical spectroscopy into a cartoon scenario: one shell is expelled on Day 0 with a velocity of $\sim$1,900 km s$^{-1}$ and a second is expelled later on with a velocity of 3,000 km s$^{-1}$. The bright hard X-ray component appears in the \emph{Swift} data on Day 117, and rises steeply until it reaches maximum on Day 142 and plateaus until Day $\sim$206, after which the flux gradually fades.  In our model, the period of the X-ray rise and plateau corresponds to the time it takes for the shock to plow through the first ejection \citep{O'Brien94}; by Day 206, the first ejection would be completely shocked. Using these measured velocities and requiring that the outermost edges of the first and second ejections catch up with one another around Day 206, we find that the second ejection was released on Day 75 (in reasonable agreement with the increase of H$\beta$ expansion velocity and the rise of radio flux density; Paper I, \citealt{Surina14}).

In addition, the hard X-ray component shows a remarkably strong \ion{O}{8} Ly$\alpha$ line in the \emph{Suzaku} spectrum, implying that the shocked gas is over-abundant in oxygen by a factor of $\sim$1--2 orders of magnitude over solar values, as is commonly observed in nova ejecta \citep[e.g.,][]{Gehrz98}. Such an unusually high oxygen abundance implies that the shocked material is nova ejecta (rather than solar-abundance material which may have been stripped from the secondary star). 
 
 
 \begin{figure}[t] 
\begin{center}
\includegraphics[height=3.4in,angle=90]{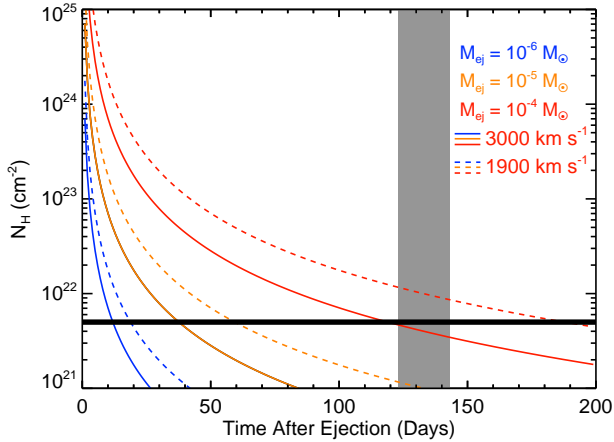}
\caption{When the column density of nova ejecta drops sufficiently (to $\sim 5 \times 10^{21}$ cm$^{-2}$; black solid line), the nuclear burning white dwarf should become visible as a super-soft X-ray source. The vertical grey region shows the plausible range of turn-on time for the super-soft phase in T~Pyx, implying an ejecta mass $\gtrsim 10^{-5}$ M$_{\odot}$. We overplot column density as a function of time for six different sets of ejecta parameters: those assuming $M_{\rm ej} = 10^{-6}$ M$_{\odot}$ (blue), $M_{\rm ej} = 10^{-5}$ M$_{\odot}$ (orange), and $M_{\rm ej} = 10^{-4}$ M$_{\odot}$ (red). Solid lines in blue, orange, and red represent ejecta with $v_{\rm max} = 3,000$ km s$^{-1}$, while dashed lines are for $v_{\rm max} = 1,900$ km s$^{-1}$.}
\label{nh}
\end{center}
\end{figure}

The neutral hydrogen column density shielding the shocked material decreases only a small amount, if at all, between Days 142 and 206, and it is not significantly higher than the ISM foreground value during this period (Figure \ref{hratio}f). Much more dramatic declines in N(H) are seen in the hard X-ray components of novae like V1974~Cyg and V382~Vel \citep{Balman98, Mukai01}. T~Pyx's small decrease in N(H) implies that the shock is not viewed through a dense absorbing screen and is not deeply embedded in the ejecta. 

We conclude that the timing, temperature, and abundances of the bright hard X-ray component are all well described by a shock within the ejecta of the 2011 outburst.

\section{A Super-soft Source Emerges Around Day 130 in the 2011 Outburst of T~Pyx} \label{sss}
We clearly observe the unveiling of the super-soft X-ray emission in the {\it Swift} observations of T~Pyx.  The super-soft phase is thought to begin when the nova ejecta have expanded enough to become optically thin to X-ray photons emanating from the still-burning shell on the surface of the white dwarf  \citep[e.g.,][]{Krautter08, Schwarz11}. We note that this super-soft turn-on is an observationally defined phenomenon; the hot white dwarf is almost certainly present at earlier times, but it is not observed because of the large absorbing column.

Our analysis of both the {\it Swift} and {\it Suzaku} spectra of T~Pyx show clear evidence for a blackbody component with temperature of 35-45 eV ($[4-5] \times 10^{5}$ K).  The average bolometric luminosity of this blackbody between Days 143 and 150 is $(1.3^{+8.0}_{-0.9}) \times 10^{38}$ erg~s$^{-1}$, near the Eddington luminosity and typical of a still shell-burning white dwarf (we note that blackbody model fits to super-soft spectra are known to underestimate the temperature and overestimate the luminosity of the white dwarf photosphere, so the uncertainties are likely larger than the statistical errors from the model fit; \citealt{Osborne11b}). The luminosity of the blackbody is variable and appears to decline over the period Days 143--206.  Similar variations have been observed in other novae, such as V2491 Cyg, where the bolometric luminosity during the super-soft phase did not appear to be constant \citep{Page10}.

The observed turn-on time of the super-soft source phase in novae has been identified as an important diagnostic of the ejecta mass, since more massive ejecta will take longer to thin out as they expand, eventually reaching a column density where soft X-rays produced at the white dwarf photosphere can be transmitted through the ejecta. \citet{Schwarz11} and \citet{Henze11} explore the time elapsed between thermonuclear runaway and the observed turn-on of the super-soft source for large samples of novae. They find that super-soft turn-on time is correlated with the $M_{\rm ej}/v_{\rm max}$. Figure \ref{nh} shows the expected temporal evolution of the column density for shells ejected at Day 0 with a range of masses and two possible expansion velocities (chosen based on the H$\beta$ line profiles of T~Pyx at Days 2 and 69 respectively; Paper I).  We assume that the shells have an $r^{-2}$ density profile and spherical symmetry that arise from a Hubble flow ejection with $v_{\rm min} / v_{\rm max} = 0.2$ (see Paper I for details, also \citealt{Seaquist_Bode08}). We estimate that $5 \times 10^{21}$ cm$^{-2}$ is the maximum column density for which we could detect a super-soft component with temperature kT $<$ 50 eV (horizontal black line).

The super-soft source is definitely observed on Day 143, the first observation after the solar gap.  A careful examination of the data obtained on Days 120 and 123 constrain the earliest turn-on time to Day 123, when equal numbers of photons are detected above and below 0.5 keV, and a hint of an additional soft component can be seen in the spectrum. In Figure \ref{nh}, we find that a turn-on time of 123 days corresponds to an ejected of mass of $\gtrsim$few $\times 10^{-5}$ M$_{\odot}$, assuming that the ejecta have been expanding since Day 0.  If there is a two-month long delay in the ejection of the bulk of the mass, as suggested by the following section, the ejecta mass implied by the turn-on time can be slightly lower,  but still $\gtrsim$10$^{-5}$ M$_{\odot}$. These mass estimates, while larger than traditionally assumed for T~Pyx because of its short recurrence time (e.g., \citealt{Contini97, Schaefer10a}), are compatible with estimates obtained from analysis of the 1966 outburst \citep{Selvelli08}, radio light curves (Paper I), and the decrease of the binary period after the 2011 outburst \citep{Patterson13}.

\begin{figure*}[t]
\begin{center}
\vspace{-1.1in}
\hspace{0.1in}
\includegraphics[height=6.8in,angle=270]{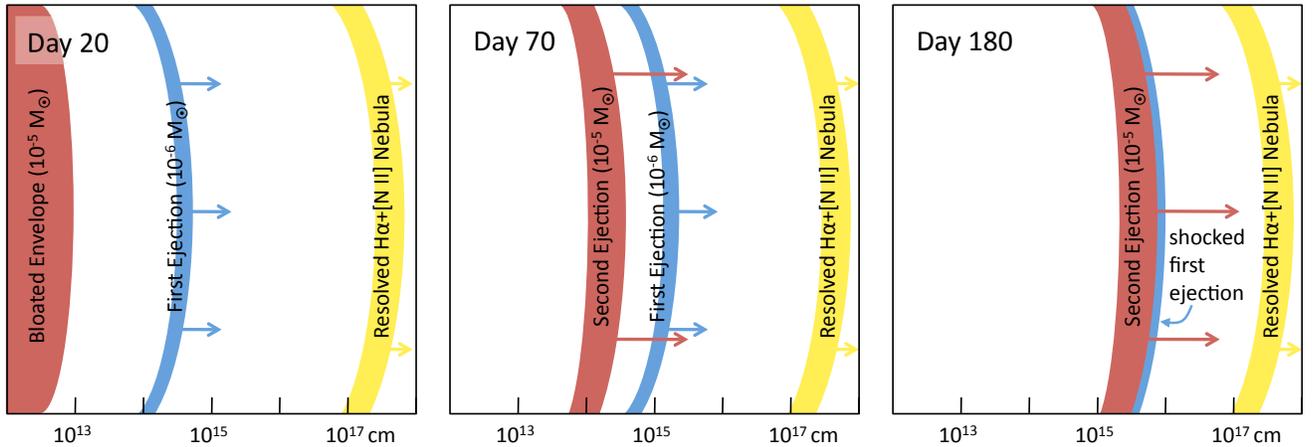}
\vspace{-1.35in}
\caption{A schematic illustration of mass ejection and interaction in the 2011 outburst of T~Pyx, integrating constraints from optical, radio, and X-ray data. The yellow, blue and red shells represent the large scale remnant in T Pyx, the first, low-mass ejection, and the later, high-mass shell, respectively.  The size of the arrows indicates the expansion velocity of each shell.  The three panels show key components in the system at Day 20 (left), Day 70 (center), and Day 180 (right); see Section \ref{multi} for more details.}
\label{cartoon}
\end{center}
\end{figure*}

The blackbody temperature associated with the super-soft component is much lower than in other recurrent novae, implying a white dwarf in T~Pyx which is significantly below the Chandrasekhar mass. More massive white dwarfs have smaller radii and surface areas, and so have higher temperature photospheres once they become super-soft sources \citep{Sala05, Wolf13}.  The recurrent novae RS~Oph and U~Sco were both observed to have high blackbody temperatures during their super-soft phases (60--90 eV; \citealt{Osborne11b,Orio13}), as was the classical nova V2491 Cyg \citep{Page10}.  T~Pyx is strikingly different, with a blackbody temperature of 30--50 eV, just half of what is measured in other recurrent novae (see also \citealt{Tofflemire13}).  Both the lower peak temperature and massive ejected shell are compatible with a nova outburst on a lower mass white dwarf ($\sim$1 M$_{\odot}$; \citealt{Wolf13}).  

The turn-off time of the super-soft source is also theoretically predicted to inversely scale with the white-dwarf mass. The turn-off of the super-soft source refers to an abrupt decline in the blackbody luminosity, and likely marks the cessation of nuclear burning on the white dwarf surface, due to the exhaustion of the reservoir of hydrogen remaining after nova outburst. On a more massive white dwarf, the nuclear-burning luminosity is higher and the reservoir of hydrogen is less massive, implying a shorter duration for the super-soft phase \citep[e.g.,][]{Sala05}.

We find evidence that the super-soft emission in T~Pyx began to turn off around Day 180 (Section 4.2). This inference from the \emph{Swift} data is in agreement with the report of a very low-luminosity blackbody component in the {\it Chandra} spectrum obtained on Day 210 \citep{Tofflemire13}. \citet{Wolf13} predict a stable shell-burning super-soft phase for $\gtrsim$1,000 days following a nova on a 1.0 M$_{\odot}$ white dwarf; a super-soft duration of $\sim$200 days is more consistent with a nova on a 1.15 M$_{\odot}$ white dwarf. The rapid super-soft turn-off time in T~Pyx therefore implies a somewhat more massive white dwarf, potentially at odds with the late observed turn-on time and the low effective temperature in the super-soft phase \citep[e.g.,][]{Henze14}. However, as we show in the next section, mass loss from T~Pyx is peculiar and complex; discrepancies in white dwarf mass constraints might be resolved if a larger fraction of the hydrogen envelope is expelled from T~Pyx than theoretically predicted.

\section{X-ray, Optical, and Radio Evidence for a Stalled Ejection in T~Pyx} \label{multi}

In Section \ref{intsh}, we found that the hard X-rays in T~Pyx are consistent with an interaction between two discrete shells: one which is triggered at Day 0 and expands at 1,900 km s$^{-1}$, and the other which begins expanding at 3,000 km s$^{-1}$ on Day $\sim$75. This picture is also consistent with peculiarities seen at radio and optical wavelengths, as we discuss here.

The shapes of the radio and X-ray light curves are remarkably similar (Figure \ref{rxo}), which at first glance is surprising, as they are likely emitted by fundamentally different processes and regions. Both essentially show non-detections for the first couple months, then rise steeply (Figure \ref{rxo}).  However, the quasi-coordinated rise is not a coincidence if the increases in both the radio and X-ray regimes are tied to the expansion of the ejecta. The radio flux rises as the optically-thick thermal ejecta expand and the emitting area increases in size. The expansion of the ejecta also leads to a drop in the column density shielding the hot white dwarf, revealing its super-soft X-ray emission.  The hard X-ray emission is produced when fast-moving ejecta plow into slower-moving material.  Based on arguments in Section 5, this material also appears to originate in the nova ejecta.

A detailed look at the two light curves reveals that the rises in the radio and X-ray are \emph{not} perfectly synchronized. The first detection of the main radio rise was made on Day 67, while the first significant detection of the main X-ray rise did not take place until 1.5 months later, Day 117. The lag between the hard X-ray and radio light curves cannot be explained if they share a common source, with the radio emission originating in the same shocked plasma as the X-ray. In addition, the radio emission during Days 117--164 is far too bright and optically thick, and the temperature of the hard X-ray component is far too hot, to share a common source in the shocked gas (see Appendix).

In Paper I, we showed that the peculiar radio light curve can be explained as either 1) a very cold ejection which is suddenly heated (requiring an increase in temperature by a factor of 25 in $\sim$19 days) or 2) a delay in the expulsion of the bulk of the mass by $\sim$60 days after the beginning of the optical rise. While both scenarios are consistent with the super-soft X-ray evolution of T~Pyx, cooling and subsequent re-heating of the ejecta provides no natural explanation for the presence of a late-onset hard X-ray component that appears to be consistent with shock emission. Even when the white dwarf's photoionizing radiation can reach the outer ejecta and heat them, the radiation is too soft to result in X-rays with E $>$ 1 keV.  In contrast, the second scenario (a delay in ejecting the bulk of the mass from the binary; Figure \ref{cartoon}) can self-consistently explain both the soft and hard X-ray components, along with the radio and optical evolution.
 
Our proposed multistage ejection scenario is illustrated in Figure 5.  First, a shell constituting a small fraction ($\lesssim$10\%) of the total envelope mass (shown in {\it blue} in Figure 5) is promptly expelled at 1,900 km s$^{-1}$ around the time of optical rise. Meanwhile, the remaining bulk of the accreted envelope (shown in {\it red}) puffs up into a quasi-hydrostatic envelope surrounding the binary for $\sim$60--70 days. This stalled, massive shell, orbiting the central binary system as a common envelope-like structure with radius $\lesssim$10$^{14}$ cm, explains the bright, roughly-constant plateau in the optical light curve and the early-time radio non-detections (\citealt{Hachisu04}; Paper I). It is interesting to note that models of radio light curves imply that a low-mass ejection expelled on Day 0 might account for the sole radio detection at early times (measured on Day 17 and followed by a suite of non-detections on Days 31--48; Paper I), which is also coincident with the very early and short-lived detection of hard X-rays.

Optical spectra reveal that T~Pyx originally resembled a He/N nova but switched to a \ion{Fe}{2} nova sometime between Day 2 and 10 (\citealt{Ederoclite13, Williams13}; see \citealt{Williams92} and \citealt{Williams12} for more interpretation of such spectral classifications). A wind with a low mass-loss rate but high velocity ($\dot{M}_{w} \approx 10^{-8}$ M$_{\odot}$ yr$^{-1}$, $v_{w} \approx$ 2,000 km s$^{-1}$) emanating from the puffed-up envelope may sustain the observed high-velocity wings on optical emission lines during Days $\sim$2--60 \citep{Surina14} while remaining optically thin and below the detection limits in the radio on Days 31--48. 

On Day 65, the optical spectrum of T~Pyx was observed to transition {\it back} to a He/N classification \citep{Ederoclite13}, implying that the bulk of the ejected mass was now being lost in a discrete shell-like event (central panel of Figure \ref{cartoon}). Around this time, the optical light curve begins to decline steeply and monotonically; we interpret this decline as the optical photosphere finally receding through the ejecta (Figure \ref{rxo}). In addition, the H$\alpha$ line flux begins to systematically decrease \citep{Surina14}, implying a drop in the density of the envelope. It is also around Day 65 that we see the radio light curve begin to rise, implying that the envelope is now growing in size (Paper I). And lastly, in our internal shock model for the hard X-ray emission from T~Pyx, we calculated that the faster moving ejection was likely expelled around Day 75 (Section \ref{intsh}; not precisely matching transitions at other wavelengths, but reasonably close given the uncertainties involved in our interpretation of the hard X-ray light curve). 
The agreement in timing from optical, radio, and X-ray tracers is remarkable; all imply that the bulk of the ejecta mass was not expelled from the environs of T~Pyx until $\sim$2 months after the thermonuclear runaway.

The physical cause of this stalled expansion remains unclear. Other novae have shown similar plateaus around optical maximum, with the duration of the plateau widely varying between sources, from a few to hundreds of days \citep[e.g.,][]{Kato11}. However, most are not graced with rich multi-wavelength data sets like the community obtained for T~Pyx in 2011, so these plateaus have proved difficult to interpret. \citet{Friedjung92} noted that during the plateau in HR~Del, the optical spectrum showed little evidence of sustained strong mass loss as seen in other novae; it was in fact more consistent with an almost stationary atmosphere.  The author proposed that the nova outburst in this system was initially not strong enough to eject the accreted envelope, and instead settled into a quasi-static configuration with little mass loss.  No explanation was offered, however, for the transition to the subsequent phase of significant mass loss. \citet{Hachisu04} concurred that plateaus around optical maximum could be interpreted as a static phase of the outburst, with an inflated but stable white dwarf photosphere formed after the onset of the thermonuclear runaway.  

In later work, \citet{Kato11} proposed that the interaction of the companion star with the hydrostatic envelope could provide an additional source of energy, empowering the outburst to transition from a stable to a mass-losing configuration. \citet{Kato11} find that this mechanism should be most efficient in very short period binaries, like T~Pyx, where the companion becomes deeply embedded in the nova envelope. In addition, \citet{Livio90} and \citet{Lloyd97} find that novae which undergo such a common-envelope-like phase may display ring-like morphologies in their ejecta, aligned with the orbital plane; \citet{Sokoloski13} find hints of a ring structure in the H$\alpha$+[\ion{N}{2}] nebula surrounding T~Pyx.  

A different explanation for a delayed ejection from T~Pyx is proposed by \citet{Williams12}; he suggests that a significant fraction of the mass loss may have its source in the binary companion star, rather than the white dwarf. If the nova explosion sweeps up or ablates material from the companion star, some of the peculiarities of T~Pyx, like its hyper-hybrid spectral classification, could be explained.   Detailed modeling of the interaction of the nova event with the companion star is required to assess the energetics, mass budget, and timescale of the interaction and evaluate if mass loss from the companion star might help explain the 2011 outburst of T~Pyx.

Even more speculatively, peculiarities like multiple and delayed ejections might be explained if the nuclear burning rate on the white dwarf surface fluctuates with time. Novae show an astounding diversity of light curves \citep{Strope10}, some of which are difficult to explain with a single thermonuclear runaway. Indeed, nuclear-driven X-ray bursts on neutron stars sometimes display multiple thermonuclear runaways in the course of a single outburst (e.g., \citealt{intZand03}). While multiple thermonuclear runaways are not theoretically predicted in a single nova outburst, they can not be ruled out by current observations.

Considering all the available data in tandem, we favor a delayed ejection as the primary explanation for the multi-wavelength evolution of T~Pyx, because this scenario can self-consistently explain the late and steep rise in the radio and X-ray light curves, the slow evolution of the optical light curve, and the unusually complex optical spectral evolution of T~Pyx.  The physical mechanism driving this late-time mass ejection remains unclear.


\section{Conclusions}

In our analysis of \emph{Swift} and \emph{Suzaku} data covering the 2011 outburst of the recurrent nova T~Pyx, we detect two distinct components of X-ray emission: a super-soft component associated with the photosphere of the nuclear-burning white dwarf, and a hard component associated with shocked thermal gas.

The super-soft X-ray component becomes detectable between Days 123--143, implying an ejection which is surprisingly massive for a recurrent nova ($\gtrsim 10^{-5}$ M$_{\odot}$), but consistent with other recent constraints for T~Pyx \citep[Paper I]{Selvelli08, Patterson13}. In addition, the temperature of the super-soft source is relatively cool (30--50 eV), implying that the white dwarf in T~Pyx is significantly below the Chandrasekhar mass. T~Pyx therefore inhabits a poorly-explored corner of nova parameter space, where its unusually high accretion rate, rather than an unusually massive white dwarf, drives it to have a short nova recurrence time.

A hard X-ray component is also detected in all epochs with sufficient signal-to-noise for spectral analysis (Days 142--206). Hard X-rays are relatively common in novae, and are postulated to originate in shocks internal to the nova ejecta \citep{O'Brien94, Krautter08}. The hard X-rays in T~Pyx, however, are accompanied by an unusually strong \ion{O}{8} Ly$\alpha$ emission line, which can only be fit if the thermal plasma has a highly super-solar abundance of oxygen. Such high oxygen abundances in the nova ejecta require the dredge-up of significant amounts of white dwarf materials during the nova event (contrary to some models of recurrent novae; e.g., \citealt{Starrfield12}). Similar conclusions can be reached from analysis of \emph{Chandra} LETG spectra, which find a factor of $\sim$15 overabundance of nitrogen in the ejecta, relative to solar (consistent with enrichment by CNO-processed white dwarf material; \citealt{Tofflemire13}).

In addition, we note the appearance of a faint, hard X-ray component at early times with an unknown origin. This component was only detectable by \emph{Swift}/XRT in time-averaged spectra spanning Days 14--20, and its low signal-to-noise precludes detailed spectral analysis. However, we note a single coincident radio detection with the VLA at 33 GHz on Day 17 (Paper I). The nature of this component remains a mystery, although it may be linked to the first low-mass ejection from T~Pyx on Day 0 (Figure \ref{cartoon}).

As measured from optical emission line profiles, the expansion velocity of ejecta in T~Pyx increases by 50\% during the first two months of outburst (Paper I, \citealt{Surina14}); we find that this variation, and subsequent interaction within the ejecta, can naturally explain the temperature and light curve of the hard X-ray component. In Section \ref{multi}, we present a cartoon picture that can self-consistently explain the soft and hard X-ray evolution, the radio light curve, the optical spectral evolution, and the optical light curve. A shell of material is expelled on Day 0 at 1,900 km s$^{-1}$, and a second episode of mass ejection is released on Day $\sim$65 at 3,000 km s$^{-1}$ (Figure \ref{cartoon}). While the first ejection produces clearer signatures at optical wavelengths, the second ejection accounts for the bulk of the expelled mass (as implied by the small absorbing column in the hard X-ray component and the bright but delayed maximum in the radio light curve). We do not yet understand the physical mechanism that leads to the bulk of the nova mass stalling out and lingering around the binary for two months before finally being expelled, but T~Pyx now joins a significant and growing sample of novae which show evidence for complex, multi-phase mass ejection (see \citealt{Lynch08, Krauss11, Williams12} for other examples).
 
\acknowledgements
We are grateful to R.~Williams, A.~Ederoclite, M.~Bode, R.~Smith, and U.~Munari for useful conversations. We thank the {\it Suzaku} mission for the generous allocation of target-of-opportunity time to observe T~Pyx.  We also thank Neil Gehrels and the \emph{Swift} mission team for their support of the target-of-opportunity program for this nova. We acknowledge with thanks the variable star observations from the AAVSO International Database contributed by observers worldwide and used in this research. This work made use of the HEASARC archive, data supplied by the UK Swift Science Data Centre at the University of Leicester, and observations obtained with the \emph{Suzaku} satellite, a collaborative mission between the space agencies of Japan (JAXA) and the USA (NASA). The National Radio Astronomy Observatory is a facility of the National Science Foundation operated under cooperative agreement by Associated Universities, Inc.  L.~Chomiuk is a Jansky Fellow of the National Radio Astronomy Observatory.  J.~Osborne and K.~Page acknowledge the support of the UK Space Agency. J.~L.~Sokoloski and J.~Weston acknowledge support from NSF award AST-1211778.

{\it Facilities:} \facility{Swift}, \facility{Suzaku}, \facility{VLA}, \facility{AAVSO}

\bibliography{tpyx.bib}

\appendix
\section{Predicting Radio Luminosity of the Hard X-ray Emitting Gas}

Here we investigate if the same thermal gas which emits hard X-rays in T~Pyx can account for a significant portion of the observed radio flux. Take for example Day 156 (2011 Sep 17), a date during the optically thick rise of the radio light curve when both VLA radio and \emph{Swift} X-ray observations were obtained. The X-ray observations imply a volume emission measure of $2.36 \times 10^{57}$ cm$^{-3}$ and a temperature of 1.2 keV ($1.4 \times 10^{7}$ K).

Assuming a spherical shell with radius of 1,900 km s$^{-1} \times$ 156 Days = $2.6 \times 10^{15}$ cm and thickness of 10\% the radius, this $EM_V$ corresponds to a number density of $n = 3 \times 10^{5}$ cm$^{-3}$. The path-length emission measure, $EM = \int n^2 dl$, determines the radio optical depth and therefore radio luminosity; here it is $EM = 2 \times 10^{7}$ cm$^{-6}$ pc. The optical depth of radio emission is:
\begin{equation}
\tau_{\nu} = 0.08235\ \left({{EM}\over{\rm cm^{-6}\, pc}}\right)\ \left({{T_{\rm sh}}\over{\rm K}}\right)^{-1.35}\ \left({{\nu}\over{\rm GHz}}\right)^{-2.1}
\end{equation}
\citep{Seaquist_Bode08}. Therefore, the radio optical depth of the shocked gas is $\tau_{\nu} = 2 \times 10^{-6}$ at 12.6 GHz, much less than unity and implying that the radio emission is completely optically thin. The radio flux density expected from this shocked gas is:
\begin{equation}
\left({{S_{\nu}}\over{\rm mJy}}\right) =  1.39 \times 10^{8} \left({{T_{\rm sh}}\over{\rm K}}\right)\ \left({{\nu}\over{\rm GHz}}\right)^{2}\ \left({{R_{\rm sh}}\over{D}}\right)^2\ (1 - e^{-\tau_{\nu}})
\end{equation}
where $R_{\rm sh}$ is the radius of the shock and $D$ is the distance to the emitting body, both in cm. From the hard X-ray emission measured on Day 156, we therefore predict a 12.6 GHz flux density of 0.02 mJy. On that day, we instead measured a 12.6 GHz flux density from T~Pyx of  11 mJy, a factor of 550 brighter than expected from the X-ray emitting shocked gas.

Similar results hold between Days 67--164; the shocked material that produces the hard X-rays (or upper limits on the hard X-rays) is not sufficient to account for the the radio emission. The bulk of the radio emission must be coming from another source. This estimate underlines the well-known result that warm $10^4$ K gas is an efficient emitter at radio wavelengths, while the hot $>$10$^6$ K gas which efficiently emits at X-ray wavelengths is usually optically thin and relatively faint in the radio.

\end{document}